# It is not Higgs

Gunn Quznetsov

## Abstract

The basic concepts, principles and statements of the electroweak and the quark-gluon theories and the theory of gravitation are deduced from properties of the point-like events probabilities. Higgs, strings, Dark Energy and Dark Matter are not required.

> «… Они пилили гири …
>   -- Что такое! -- сказал вдруг Балаганов, переставая работать. -- Три часа уже пилю, а оно все еще не золотое.
>   Паниковский не ответил. Он уже все понял и последние полчаса водил ножовкой только для виду.
>  -- Ну-с, попилим еще! - бодро сказал рыжеволосый Шура.
>   -- Конечно, надо пилить, - заметил Паниковский, стараясь оттянуть страшный час расплаты…».
>
> Илья Ильф и Евгений Петров, «Золотой Теленок», <http://www.lib.ru/ILFPETROV/telenok.txt> [1]

## Introduction

Let us denote:

$\vec{x} := \langle x_1, x_2, x_3 \rangle$;

$x_0 := ct$ with $c = 299\ 792\ 458$;

$\underline{x} := \langle x_0, \vec{x} \rangle$;

$\int d^{3+1}\underline{x} := \int dx_0 \int dx_1 \int dx_2 \int dx_3$;

$\int d^3\vec{x} := \int dx_1 \int dx_2 \int dx_3$;

$1_2 := \begin{bmatrix} 1 & 0 \\ 0 & 1 \end{bmatrix}, 0_2 := \begin{bmatrix} 0 & 0 \\ 0 & 0 \end{bmatrix}, \beta^{[0]} := -1_4 := -\begin{bmatrix} 1_2 & 0_2 \\ 0_2 & 1_2 \end{bmatrix}$;

---

[1] "... They sawed dumb-bells ...
- What! - Said Balaganov suddenly, stopping work. - Three hours already I saw, but it still is not gold.
  Panikovsky not answered. He understood everything and drove the last half hour with a hacksaw just for show.
 - Well, saw some more! - Said cheerfully redhead Shura.
- Of course, it is necessary to saw - saw Panikovsky, trying to delay the terrible day of reckoning…".
Ilya Ilf and Yevgeny Petrov, "The Golden Calf"
http://www.lib.ru/ILFPETROV/telenok.txt

if $A$ is a $2 \times 2$-matrix then:

$$A1_4 := \begin{bmatrix} A & 0_2 \\ 0_2 & A \end{bmatrix} \text{ and } 1_4 A := \begin{bmatrix} A & 0_2 \\ 0_2 & A \end{bmatrix}$$

and if $B$ is $4 \times 4$-matrix then:

$$A + B := A1_4 + B, AB := A1_4 B;$$

the Pauli matrices:

$$\sigma_1 := \begin{bmatrix} 0 & 1 \\ 1 & 0 \end{bmatrix}, \sigma_2 := \begin{bmatrix} 0 & -i \\ i & 0 \end{bmatrix}, \sigma_3 := \begin{bmatrix} 1 & 0 \\ 0 & -1 \end{bmatrix}.$$

In 1963 American physicist Sheldon Glashow [1] proposed that the weak nuclear force and electricity and magnetism could arise from a partially unified electroweak theory. But "… there is major problem: all the fermions and gauge bosons are massless, while experiment shows otherwise.

Why not just add in mass terms explicitly? That will not work, since the associated terms break SU(2) or gauge invariantnes. For fermions, the mass term should be $m\bar{\psi}\psi$

$$m\bar{\psi}\psi = m\bar{\psi}(P_L + P_R)\psi$$
$$= m\bar{\psi}P_L P_L \psi + m\bar{\psi}P_R P_R \psi$$
$$= m(\bar{\psi}_R \psi_L + \bar{\psi}_L \psi_R).$$

However, the left-handed fermion are put into SU(2) doublets and the right-handed ones into SU(2) singlets, so $\bar{\psi}_R \psi_L$ and $\bar{\psi}_L \psi_R$ are not SU(2) singlets and would not give an SU(2) invariant Lagrangian.

Similarly, the expected mass terms for the gauge bosons,

$$\frac{1}{2} m_B^2 B^\mu B_\mu$$

plus similar terms for other, are clearly not invariant under gauge transformations $B_\mu \to B'_\mu = B_\mu - \frac{\partial_\mu \chi}{g}$. The only direct way to preserve the gauge invariance and SU(2) invariance of Lagrangian is to set $m = 0$ for all quarks, leptons and gauge bosons…. There is a way to solve this problem, called the Higgs mechanism." [2]

No. The Dirac Lagrangian for a free fermion can have of the following form:

$$\mathcal{L}_f := \bar{\psi}(\beta^{[0]}\partial_0 + \beta^{[1]}\partial_1 + \beta^{[2]}\partial_2 + \beta^{[3]}\partial_3 + m\gamma^{[0]})\psi.$$

Here matrices $\beta^{[1]}$, $\beta^{[2]}$, $\beta^{[3]}$, and $\gamma^{[0]}$ anticommute among themselves.
Indeed, this Lagrangian is not invariant under the SU(2) transformation. But it is beautiful and truncating its mass term is not good idea.

But it turns out that there is a fifth matrix $\beta^{[4]}$ anticommuting with these four matrices. And the term with this matrix should be added to this Lagrangian mass term:

$$\mathcal{L}_f' := \overline{\psi}(\beta^{[0]}\partial_0 + \beta^{[1]}\partial_1 + \beta^{[2]}\partial_2 + \beta^{[3]}\partial_3 + m_1\gamma^{[0]} + +m_2\beta^{[4]})\psi.$$

where $\sqrt{m_1^2 + m_2^2} =: m$.

Let $U(\alpha)$ be any SU(2)-matrix with parameter $\alpha$, let $\mathcal{U}$ be the space in which $U(\alpha)$ acts. In such case $U(\alpha)$ divides the space $\mathcal{U}$ into two orthogonal subspaces $\mathcal{U}_o$ and $\mathcal{U}_x$ such that for every element $\psi$ of $\mathcal{U}$ there exists an elements $\psi_o$ of $\mathcal{U}_o$ and an element $\psi_x$ of $\mathcal{U}_x$ which fulfill the following conditions:

$$\psi = \psi_o + \psi_x,$$

$$\overline{\psi}_x U^\dagger(\alpha)(\beta^{[0]}\partial_0 + \beta^{[1]}\partial_1 + \beta^{[2]}\partial_2 + \beta^{[3]}\partial_3 + m_1\gamma^{[0]} + +m_2\beta^{[4]})U(\alpha)\psi_x =$$

$$= \overline{\psi}_x(\beta^{[0]}\partial_0 + \beta^{[1]}\partial_1 + \beta^{[2]}\partial_2 + \beta^{[3]}\partial_3 + (m_1\cos\alpha - m_2\sin\alpha)\gamma^{[0]}$$
$$+ (m_2\cos\alpha + m_1\sin\alpha)\beta^{[4]})\psi_x,$$

$$\overline{\psi}_o U^\dagger(\alpha)(\beta^{[0]}\partial_0 + \beta^{[1]}\partial_1 + \beta^{[2]}\partial_2 + \beta^{[3]}\partial_3 + m_1\gamma^{[0]} + +m_2\beta^{[4]})U(\alpha)\psi_o =$$

$$= \overline{\psi}_o(\beta^{[0]}\partial_0 + \beta^{[1]}\partial_1 + \beta^{[2]}\partial_2 + \beta^{[3]}\partial_3 + (m_1\cos\alpha + m_2\sin\alpha)\gamma^{[0]}$$
$$+ (m_2\cos\alpha - m_1\sin\alpha)\beta^{[4]})\psi_o.$$

In either case, $m$ does not change.

What is concerning the gauge bosons $W_\mu$ and $Z_\mu$: although field $F_{\mu\nu}$

$$(F_{\mu\nu} = \partial_\mu W_\nu - \partial_\nu W_\mu - i\frac{g_2}{2}(W_\mu W_\nu - W_\nu W_\mu))$$

is massless, it turns out that its components obey the equation which is similar to the Klein-Gordon equation with nonzero mass[2].

Therefore, Higgs is neither required to obtain masses fermions, nor to obtain masses of bosons. Moreover, the Tevatron[3] has left little hope for the experimental finding the Higgs.

In this article I consider the events, each of which can bound to a certain point in space-time. Such events are called pointlike events [3]. Combinations (sums, products, supplements) of such events are events, called physical events.

The probability density of pointlike events in space-time is invariant under Lorentz transformations. But probability density of such events in space at a certain instant of time is not invariant under these transformations. I consider the pointlike events for which density of probability in space at some instant of time is the null component of a 3+1-vector function which is transformed by the Lorentz formulas, I call these probabilities *the traceable probabilities*.

The set $\check{C}$ of complex $n \times n$ matrices is called a Clifford set of range $n$, if here the following conditions are fulfilled:

If $\alpha_k \in \check{C}$ and $\alpha_s \in \check{C}$ then $\alpha_k\alpha_s + \alpha_s\alpha_k = 2\delta_{k,s}$;
if $\alpha_k\alpha_s + \alpha_s\alpha_k = 2\delta_{k,s}$ for all $\check{C}$'s elements $\alpha_s$ then $\alpha_k \in \check{C}$.

If $n = 4$ then either Clifford set contains three elements (*Clifford triplet*) or it contains five elements (*Clifford pentad*) [4].

---
[2] see part *Electroweak equations* of this paper.
[3] For example: http://www-d0.fnal.gov/Run2Physics/WWW/results/prelim/HIGGS/H88/H88.pdf

Here exist only six Clifford pentads: one, which I call *light pentad* $\beta$:

$$\beta^{[1]} := \begin{bmatrix} \sigma_1 & 0_2 \\ 0_2 & -\sigma_1 \end{bmatrix}, \beta^{[2]} := \begin{bmatrix} \sigma_2 & 0_2 \\ 0_2 & -\sigma_2 \end{bmatrix}, \beta^{[3]} := \begin{bmatrix} \sigma_3 & 0_2 \\ 0_2 & -\sigma_3 \end{bmatrix}, \quad (2)$$

$$\gamma^{[0]} := \begin{bmatrix} 0_2 & 1_2 \\ 1_2 & 0_2 \end{bmatrix}, \quad (3)$$

$$\beta^{[4]} := \begin{bmatrix} 0_2 & i1_2 \\ -i1_2 & 0_2 \end{bmatrix}. \quad (4)$$

Three *chromatic pentads*:
  *the red pentad* $\zeta$:

$$\zeta^{[1]} := \begin{bmatrix} -\sigma_1 & 0_2 \\ 0_2 & \sigma_1 \end{bmatrix}, \zeta^{[2]} := \begin{bmatrix} \sigma_2 & 0_2 \\ 0_2 & \sigma_2 \end{bmatrix}, \zeta^{[3]} := \begin{bmatrix} -\sigma_3 & 0_2 \\ 0_2 & -\sigma_3 \end{bmatrix},$$

$$\gamma^{[0]}_\zeta := \begin{bmatrix} 0_2 & -\sigma_1 \\ -\sigma_1 & 0_2 \end{bmatrix}, \zeta^{[4]} := i\begin{bmatrix} 0_2 & \sigma_1 \\ -\sigma_1 & 0_2 \end{bmatrix};$$

*the green pentad* $\eta$:

$$\eta^{[1]} := \begin{bmatrix} -\sigma_1 & 0_2 \\ 0_2 & -\sigma_1 \end{bmatrix}, \eta^{[2]} := \begin{bmatrix} -\sigma_2 & 0_2 \\ 0_2 & \sigma_2 \end{bmatrix}, \eta^{[3]} := \begin{bmatrix} \sigma_3 & 0_2 \\ 0_2 & -\sigma_3 \end{bmatrix},$$

$$\eta^{[0]}_\zeta := \begin{bmatrix} 0_2 & -\sigma_2 \\ -\sigma_2 & 0_2 \end{bmatrix}, \eta^{[4]} := i\begin{bmatrix} 0_2 & \sigma_2 \\ -\sigma_2 & 0_2 \end{bmatrix};$$

 *the blue pentad* $\theta$:

$$\theta^{[1]} := \begin{bmatrix} \sigma_1 & 0_2 \\ 0_2 & -\sigma_1 \end{bmatrix}, \theta^{[2]} := \begin{bmatrix} \sigma_2 & 0_2 \\ 0_2 & -\sigma_2 \end{bmatrix}, \theta^{[3]} := \begin{bmatrix} -\sigma_3 & 0_2 \\ 0_2 & \sigma_3 \end{bmatrix},$$

$$\theta^{[0]}_\zeta := \begin{bmatrix} 0_2 & -\sigma_3 \\ -\sigma_3 & 0_2 \end{bmatrix}, \theta^{[4]} := i\begin{bmatrix} 0_2 & \sigma_3 \\ -\sigma_3 & 0_2 \end{bmatrix};$$

two *gustatory pentads*:
*the sweet pentad* $\underline{\Delta}$:

$$\underline{\Delta}^{[1]} := \begin{bmatrix} 0_2 & -\sigma_1 \\ -\sigma_1 & 0_2 \end{bmatrix}, \quad \underline{\Delta}^{[2]} := \begin{bmatrix} 0_2 & -\sigma_2 \\ -\sigma_2 & 0_2 \end{bmatrix}, \underline{\Delta}^{[3]} := \begin{bmatrix} 0_2 & -\sigma_3 \\ -\sigma_3 & 0_2 \end{bmatrix},$$

$$\underline{\Delta}^{[0]} := \begin{bmatrix} -1_2 & 0_2 \\ 0_2 & 1_2 \end{bmatrix}, \underline{\Delta}^{[4]} := i\begin{bmatrix} 0_2 & 1_2 \\ -1_2 & 0_2 \end{bmatrix};$$

*the bitter pentad* $\underline{\Gamma}$:

$$\underline{\Gamma}^{[1]} := i\begin{bmatrix} 0_2 & -\sigma_1 \\ \sigma_1 & 0_2 \end{bmatrix}, \quad \underline{\Gamma}^{[2]} := i\begin{bmatrix} 0_2 & -\sigma_2 \\ \sigma_2 & 0_2 \end{bmatrix}, \underline{\Gamma}^{[3]} := i\begin{bmatrix} 0_2 & -\sigma_3 \\ \sigma_3 & 0_2 \end{bmatrix},$$

$$\underline{\Gamma}^{[0]} := \begin{bmatrix} -1_2 & 0_2 \\ 0_2 & 1_2 \end{bmatrix}, \underline{\Gamma}^{[4]} := \begin{bmatrix} 0_2 & 1_2 \\ 1_2 & 0_2 \end{bmatrix}.$$

The light pentad contains three matrices corresponding to the coordinates of 3-dimensional space, and two matrices relevant to mass terms - one for the lepton and one for the neutrino of this lepton.

Each chromatic pentad also contains three matrices corresponding to three coordinates and two mass matrices - one for top quark and another - for bottom quark.

Each gustatory pentad contains one coordinate matrix and two pairs of mass matrices --- these pentads are not needed yet.

It is proven [5] that any square-integrable 4x1-matrix function with bounded domain (Planck's function) obeys some generalization of Dirac's equation with additional gauge members. This generalization is the sum of products of the coordinate matrices of the light pentad and covariant derivatives of the corresponding coordinates plus product of all the eight mass matrices (two of light and six of chromatic) and the corresponding mass numbers.

If this equation does not contain chromatic mass numbers then we obtain Dirac's equation for leptons with gauge members which are similar to electroweak fields. This equation is invariant under electroweak transformations. The Klein-Gordon type equation with nonzero mass is obtained for gauge fields $W$ and $Z$.

If this equation does not contain lepton's and neutrino's mass then we obtain the Dirac's equation with gauge members similar to eight gluon's fields. And oscillations of chromatic states of this equation bend space-time. This bend gives rise to the effects of red shift, confinement, the accelerated expansion of the space and asymptotic freedom, and Newtonian gravity turns out to be a continuation of sub nucleonic forces.

If probability of a pointlike event is limited in space-time then density of that probability at certain instant of time is represented as a square of 4x1 complex matrix Planck's function which obeys an equation of Dirac's type.

## Pointlike events

Events which are expressed by sentences of shape "In time instant $x_0$ event $A$ has coordinates $\vec{x}$" is written as the following: "$A(x_0, \vec{x})$". Such events are called *pointlike events*. In that case *a physical event* is ensemble of pointlike events.

Moreover, denote:

$$A(D) := \left(A(x_0, \vec{x}) \& \langle x_0, \vec{x} \rangle \in D\right).$$

Let $\dot{P}$ be a classical probability function.

Let $A$ be a pointlike event, and let $p_A(\underline{x})$ be such function that for every domain $D$: if $D \subseteq R^{3+1}$ then

$$\int_{(D)} d^{3+1}\underline{x} \times p_A(\underline{x}) = \dot{P}(A(D)).$$

In that case $p_A(\underline{x})$ is called *absolute density of probability* of event $A$.

If $J$ is the Jacobean of the following transformation:

$$x_0 \to x'_0 = \frac{x_0 - \frac{v}{c} x_k}{\sqrt{1 - \left(\frac{v}{c}\right)^2}}, x_k \to x'_k = \frac{x_k - \frac{v}{c} x_0}{\sqrt{1 - \left(\frac{v}{c}\right)^2}}, x_j \to x'_j = x_j \text{ if } j \neq k \quad (5)$$

Then

$$J = \frac{\partial(x'_0, \vec{x}')}{\partial(x_0, \vec{x})} = 1.$$

Hence, absolute probability density is invariant under the Lorentz transformation.

If

$$\rho_A(x_0, \vec{x}) := \frac{p_A(x_0, \vec{x})}{\int d^3 \vec{y} \cdot p_A(x_0, \vec{y})}$$

then $\rho_A(x_0, \vec{x})$ is called *probability density* of event $A$ at the time instant $x_0$.

Under transformation (5) this function is changed as the following:

$$\rho_A(x_0, \vec{x}) \to \rho'_A(x_0, \vec{x}) = \frac{p_A(x_0, \vec{x})}{\int d^3\vec{y} \cdot p_A\left(x_0 + \frac{v}{c}(y_k - x_k), \vec{y}\right)}$$

Therefore, $\rho_A(x_0, \vec{x})$ is not invariant under these transformations.

Further I consider events $A$ for which function $\rho_A$ represents the zero component of some 3+1 - vector field $j_A$ :

$$j_A := \langle \rho_A c, j_{A,1}, j_{A,2}, j_{A,3}\rangle = \langle \rho_A c, \vec{j}_A\rangle$$

$$\vec{j}_A := \langle j_{A,1}, j_{A,2}, j_{A,3}\rangle.$$

This field $\vec{j}_A$ is called *a probability current vector* of event $A$. And if $\vec{u}_A := \vec{j}_A/\rho_A$ then $\vec{u}_A$ is called *a local velocity of probability propagation* of event $A$.

That is, there exist functions $j_{A,k}$ which are changed as the following:

$$\rho_A \to \rho_A' = \frac{\rho_A - \frac{v}{c^2} j_{A,k}}{\sqrt{1 - \left(\frac{v}{c}\right)^2}}, j_{A,k} \to j_{A,k}' = \frac{j_{A,k} - v\rho_A}{\sqrt{1 - \left(\frac{v}{c}\right)^2}}, j_{A,s} \to j_{A,s}' = j_{A,s} \text{ if } s \neq k$$

The following set of 4 complex equations with unknown $4\times 1$-matrix complex function $\varphi(\underline{x})$

$$\begin{cases} \varphi^\dagger \varphi = \rho_A, \\ \varphi^\dagger \beta^{[k]} \varphi = -j_{A,k}/c \end{cases} \quad (6)$$

has solutions for any $\rho_A$ and $\vec{j}_A$ [6].

From here since $\vec{u}_A := \vec{j}_A/\rho_A$ then

$$|\vec{u}_A|^2 = u_{A,1}^2 + u_{A,2}^2 + u_{A,3}^2 \leq c^2.$$

Let $h := 6.6260755^{-34}$ and $\underline{\Omega}$ ($\underline{\Omega} \in R^{3+1}$) be a domain such that if $\underline{x} \in \underline{\Omega}$ then $|x_r| < c\pi/h$ for $r \in \{0,1,2,3\}$.

Let $\mathfrak{R}_\underline{\Omega}$ be a set of functions such that if element $\phi(\underline{x})$ of this set: if $\underline{x} \notin \underline{\Omega}$ then $\phi(\underline{x}) = 0$.

Hence,

$$\int_{(\underline{\Omega})} d^{3+1}\underline{x} \cdot \phi(\underline{x}) = \int_{-\frac{c\pi}{h}}^{\frac{c\pi}{h}} dx_0 \int_{-\frac{c\pi}{h}}^{\frac{c\pi}{h}} dx_1 \int_{-\frac{c\pi}{h}}^{\frac{c\pi}{h}} dx_2 \int_{-\frac{c\pi}{h}}^{\frac{c\pi}{h}} dx_3 \cdot \phi(\underline{x})$$

and for each element $\phi(\underline{x})$ of $\mathfrak{R}_\underline{\Omega}$ exists a number $J_\phi$ such that

$$J_\phi = \int_{(\underline{\Omega})} d^{3+1}\underline{x} \cdot \phi^*(\underline{x})\phi(\underline{x}).$$

Therefore, $\mathfrak{R}_\underline{\Omega}$ is an unitary space with the following scalar product:

$$\tilde{u} * \tilde{v} := \int_{(\underline{\Omega})} d^{3+1}\underline{x} \cdot \tilde{u}^*(\underline{x})\tilde{v}(\underline{x}).$$

This space has an orthonormalized basis with the following elements:

$$\varsigma_{w,\vec{p}}(t,\vec{x}) := \begin{cases} \left(\dfrac{h}{2\pi c}\right)^2 e^{ihwt} e^{-i\frac{h}{c}\vec{p}\cdot\vec{x}}, & \text{if } -\dfrac{\pi c}{h} \leq x_k \leq \dfrac{\pi c}{h}, \\ 0, & \text{otherwize.} \end{cases}$$

with $k \in \{1,2,3\}$ and with natural $w, p_1, p_2, p_3$ (here: $\vec{p} := \langle p_1, p_2, p_3 \rangle$ and

$\vec{p} \cdot \vec{x} := p_1 x_1 + p_2 x_2 + p_3 x_3$).

I call elements of the space with this basis *Planck's function*.
Let $j \in \{1,2,3,4\}$ and $k \in \{1,2,3,4\}$ and denote:

$$\sum_{\vec{k}} := \sum_{k_1=-\infty}^{\infty} \sum_{k_2=-\infty}^{\infty} \sum_{k_3=-\infty}^{\infty}.$$

Let $\varphi_j(t, \vec{x})$ be a Planck's function and let the Fourier series for $\varphi_j(t, \vec{x})$ has the following form:

$$\varphi_j(t, \vec{x}) = \sum_{w=-\infty}^{\infty} \sum_{\vec{p}} c_{j,w,\vec{p}}\, \varsigma_{w,\vec{p}}(t, \vec{x}).$$

If denote: $\varphi_{j,w,\vec{p}}(t, \vec{x}) := c_{j,w,\vec{p}}\varsigma_{w,\vec{p}}(t, \vec{x})$ the Fourier series for $\varphi_j(t, \vec{x})$ has the following form:

$$\varphi_j(t, \vec{x}) = \sum_{w=-\infty}^{\infty} \sum_{\vec{p}} \varphi_{j,w,\vec{p}}(t, \vec{x}).$$

Let $\langle t, \vec{x} \rangle$ be any space-time point.
Denote:

$$A_k := \varphi_{k,w,\vec{p}}|_{\langle t,\vec{x}\rangle}$$

the value of function $\varphi_{k,w,\vec{p}}$ in this point. And let us denote:

$$C_j := \left(\frac{1}{c}\partial_t \varphi_{j,w,\vec{p}} - \sum_{s=1}^{4}\sum_{\alpha=1}^{3} \beta_{j,s}^{[\alpha]} \partial_\alpha \varphi_{s,w,\vec{p}}\right)\bigg|_{\langle t,\vec{x}\rangle}$$

the value in this point of the following function:

$$\frac{1}{c}\partial_t \varphi_{j,w,\vec{p}} - \sum_{s=1}^{4}\sum_{\alpha=1}^{3} \beta_{j,s}^{[\alpha]} \partial_\alpha \varphi_{s,w,\vec{p}}.$$

There $A_k$ and $C_j$ are complex numbers. Hence, the following set of equations

$$\begin{cases} \sum_{k=1}^{4} z_{j,k,w,\vec{p}} A_k = C_j, \\ z_{j,k,w,\vec{p}}^* = -z_{k,j,w,\vec{p}} \end{cases}$$

is a system of 14 algebraic equations with complex unknowns $z_{j,k,w,\vec{p}}$.

This system has solution for any Planck's functions $\varphi_j(t, \vec{x})$ [7].

Therefore,
$$\frac{1}{c}\partial_t\varphi_{j,w,\vec{p}} = \sum_{k=1}^{4}\left(\sum_{\alpha=1}^{3}\beta_{j,k}^{[\alpha]}\partial_\alpha\varphi_{k,w,\vec{p}} + z_{j,k,w,\vec{p}}\varphi_{k,w,\vec{p}}\right)$$

in every point $\langle t, \vec{x}\rangle$.

Let $\kappa_{w,\vec{p}}$ be a linear operator on the linear space, spanned of the basic functions $\varsigma_{w,\vec{p}}(t,\vec{x})$, such that
$$\kappa_{w,\vec{p}}\varsigma_{w',\vec{p}'} = \begin{cases}\varsigma_{w',\vec{p}'} & \text{if } w' = w, \vec{p}'=\vec{p},\\ 0, & \text{otherwize.}\end{cases}$$

Let
$$Q_{j,k}|_{\langle t,\vec{x}\rangle} := \sum_{w,\vec{p}}\left(z_{j,k,w,\vec{p}}|_{\langle t,\vec{x}\rangle}\right)\kappa_{w,\vec{p}}$$
in every point $\langle t, \vec{x}\rangle$.

Therefore, for every Planck's function $\varphi_j$ here exists an operator $Q_{j,k}$ such that dependence of $\varphi_j$ on $t$ is described by the following differential equations:
$$\partial_t\varphi_j = c\sum_{k=1}^{4}\left(\beta_{j,k}^{[1]}\partial_1 + \beta_{j,k}^{[2]}\partial_2 + \beta_{j,k}^{[3]}\partial_3 + Q_{j,k}\right)\varphi_k \tag{7}$$

and
$$Q_{j,k}^* = \sum_{w,\vec{p}}\left(z_{j,k,w,\vec{p}}^*\right)\kappa_{w,\vec{p}} = \sum_{w,\vec{p}}\left(-z_{k,j,w,\vec{p}}\right)\kappa_{w,\vec{p}} = -Q_{k,j}^*.$$

The matrix form of formula (7) is the following:
$$\partial_t\varphi = c\left(\beta^{[1]}\partial_1 + \beta^{[2]}\partial_2 + \beta^{[3]}\partial_3 + \vec{Q}\right)\varphi$$

with
$$\vec{Q} := \begin{bmatrix} Q_{1,1} & Q_{1,2} & Q_{1,3} & Q_{1,4} \\ -Q_{1,2}^* & Q_{2,2} & Q_{2,3} & Q_{2,4} \\ -Q_{1,3}^* & -Q_{2,3}^* & Q_{3,3} & Q_{3,4} \\ -Q_{1,4}^* & -Q_{2,4}^* & -Q_{3,4}^* & Q_{4,4} \end{bmatrix}.$$

This equation can be rewritten as the following:

$$\sum_{k=0}^{3}\beta^{[k]}\left(\partial_k - i\Theta_k - i\Upsilon_k\gamma^{[5]}\right)\varphi$$
$$+ \left(iM_0\gamma^{[0]} + iM_4\beta^{[4]} - iM_{\zeta,0}\gamma_\zeta^{[0]} + iM_{\zeta,4}\zeta^{[4]} - iM_{\eta,0}\gamma_\eta^{[0]} - iM_{\eta,4}\eta^{[4]} + iM_{\theta,0}\gamma_\theta^{[0]}\right.$$
$$\left. + iM_{\theta,4}\theta^{[4]}\right)\varphi = 0 \tag{8}$$

with real $\Theta_k$, $\Upsilon_k$, $M_0$, $M_4$, $M_{\zeta,0}$, $M_{\zeta,4}$, $M_{\eta,0}$, $M_{\eta,4}$, $M_{\theta,0}$, $M_{\theta,4}$, and with
$$\gamma^{[5]} := \begin{bmatrix} 1_2 & 0_2 \\ 0_2 & -1_2 \end{bmatrix}.$$

Because $\zeta^{[k]} + \eta^{[k]} + \theta^{[k]} = -\beta^{[k]}$ then from (8):

$$\left(-(\partial_0 - i\Theta_k - i Y_k \gamma^{[5]}) + \sum_{k=1}^{3} \beta^{[k]}(\partial_k - i\Theta_k - i Y_k \gamma^{[5]}) + 2(iM_0 \gamma^{[0]} + iM_4 \beta^{[4]})\right)\varphi +$$

$$+\left(-(\partial_0 - i\Theta_k - i Y_k \gamma^{[5]}) - \sum_{k=1}^{3} \zeta^{[k]}(\partial_k - i\Theta_k - i Y_k \gamma^{[5]}) + 2\left(-iM_{\zeta,0}\gamma_\zeta^{[0]} + iM_{\zeta,4}\zeta^{[4]}\right)\right)\varphi +$$

$$+\left(-(\partial_0 - i\Theta_k - i Y_k \gamma^{[5]}) - \sum_{k=1}^{3} \eta^{[k]}(\partial_k - i\Theta_k - i Y_k \gamma^{[5]}) + 2\left(-iM_{\eta,0}\gamma_\eta^{[0]} - iM_{\eta,4}\eta^{[4]}\right)\right)\varphi +$$

$$+\left(-(\partial_0 - i\Theta_k - i Y_k \gamma^{[5]}) - \sum_{k=1}^{3} \theta^{[k]}(\partial_k - i\Theta_k - i Y_k \gamma^{[5]}) + 2\left(iM_{\theta,0}\gamma_\theta^{[0]} + iM_{\theta,4}\theta^{[4]}\right)\right)\varphi = 0.$$

I call the following part of equation (8)

$$\sum_{k=0}^{3} \beta^{[k]}(\partial_k - i\Theta_k - i Y_k \gamma^{[5]})\varphi + (iM_0 \gamma^{[0]} + iM_4 \beta^{[4]})\varphi = 0 \tag{9}$$

*a lepton moving equation*. And I call the following sum:

$$\widehat{H}_l := c \sum_{k=1}^{3} \beta^{[k]}(i\partial_k + \Theta_k + Y_k \gamma^{[5]}) - c(M_0 \gamma^{[0]} + M_4 \beta^{[4]}) \tag{10}$$

*a lepton Hamiltonian*.

## Masses

Let $j_{A,4} := c\varphi^\dagger \beta^{[4]} \varphi$ and $j_{A,5} := c\varphi^\dagger \gamma^{[0]} \varphi$ and similar to $\vec{u}_A := \vec{j}_A/\rho_A$, let $u_{A,4} := j_{A,4}/\rho_A$ and $u_{A,5} := j_{A,5}/\rho_A$.

In that case [8]:

$u_{A,1}^2 + u_{A,2}^2 + u_{A,3}^2 + u_{A,4}^2 + u_{A,5}^2 = c^2$.

Thus, only all five elements of a Clifford pentad provide an entire set of speed components and, for completeness, yet two "space" coordinates $x_4$ and $x_5$ should be added to our three $x_1, x_2, x_3$. These additional coordinates can be selected so that

$$-\frac{\pi c}{h} \leq x_4 \leq \frac{\pi c}{h} \text{ and } -\frac{\pi c}{h} \leq x_5 \leq \frac{\pi c}{h}.$$

Coordinates $x_4$ and $x_5$ are not coordinates of any events. Hence, our devices do not detect them as actual space coordinates.

Denote:

$\tilde{\varphi}(t, x_1, x_2, x_3, x_4, x_5) := \varphi(t, x_1, x_2, x_3)e^{i(x_5 M_0(t,x_1,x_2,x_3) + x_4 M_4(t,x_1,x_2,x_3))}$.

In that case equation (9) has the following shape:

$$\sum_{k=0}^{3}\beta^{[k]}\left(\mathrm{i}\partial_k+\Theta_k+\Upsilon_k\gamma^{[5]}\right)\tilde{\varphi}+\left(-\gamma^{[0]}\mathrm{i}\partial_5-\beta^{[4]}\mathrm{i}\partial_4\right)\tilde{\varphi}=0. \tag{11}$$

Because the following system of equations with unknown functions $B_\mu$ and $F_\mu$

$$\begin{cases}0.5g_1B_\mu-F_\mu=-\Theta_\mu-\Upsilon_\mu,\\ g_1B_\mu-F_\mu=-\Theta_\mu+\Upsilon_\mu\end{cases}$$

has solution for any $\mu$ and for some constant real positive number $g_1$ then equation (11) has got the following form:

$$\sum_{\mu=0}^{3}\beta^{[\mu]}\left(\mathrm{i}\partial_\mu+F_\mu+0.5g_1YB_\mu\right)\tilde{\varphi}+\left(-\gamma^{[0]}\mathrm{i}\partial_5-\beta^{[4]}\mathrm{i}\partial_4\right)\tilde{\varphi}=0. \tag{12}$$

with

$$Y:=-\begin{bmatrix}1_2 & 0_2\\ 0_2 & 2\cdot 1_2\end{bmatrix}.$$

Let $\chi(t, x_1, x_2, x_3)$ be a real function and let

$$\tilde{U}(\chi):=\begin{bmatrix}1_2 e^{\mathrm{i}\frac{\chi}{2}} & 0_2\\ 0_2 & 1_2 e^{\mathrm{i}\chi}\end{bmatrix}. \tag{13}$$

In that case equation (12) is invariant [9] under the following transformation:

$$\begin{cases}x_4\to x_4':=x_4\cos\dfrac{\chi}{2}-x_5\sin\dfrac{\chi}{2},\\ x_5\to x_5':=x_5\cos\dfrac{\chi}{2}+x_4\sin\dfrac{\chi}{2},\\ x_\mu\to x_\mu':=x_\mu\text{ for }\mu\in\{0,1,2,3\},\\ B_\mu\to B_\mu':=B_\mu-\dfrac{1}{g_1}\partial_\mu\chi,\\ F_\mu\to F_\mu':=F_\mu.\end{cases}$$

Let the following matrices

$$\varepsilon_1:=\begin{bmatrix}1\\0\\0\\0\end{bmatrix},\ \varepsilon_2:=\begin{bmatrix}0\\1\\0\\0\end{bmatrix},\ \varepsilon_3:=\begin{bmatrix}0\\0\\1\\0\end{bmatrix},\ \varepsilon_4:=\begin{bmatrix}0\\0\\0\\1\end{bmatrix}$$

form the basis in which elements of Clifford pentads have shape of (2), (3), (4) etc.

Functions of type

$$\frac{h}{2\pi c}e^{-\mathrm{i}\frac{h}{c}(sx_4+nx_5)}\varepsilon_k$$

with integer $n$ and $s$ form an orthonormalized basis of some unitary space $\Im$ with scalar product of the following shape:

$$(\tilde{\varphi}, \tilde{\chi}) := \int_{-\frac{\pi c}{h}}^{\frac{\pi c}{h}} dx_5 \int_{-\frac{\pi c}{h}}^{\frac{\pi c}{h}} dx_4 \cdot \tilde{\varphi}^\dagger \tilde{\chi}.$$

From (5):

$(\tilde{\varphi}, \tilde{\varphi}) = \rho_A,$

$(\tilde{\varphi}, \beta^{[s]} \tilde{\varphi}) = -\dfrac{j_{A,s}}{c}$

for $s \in \{1,2,3\}$.

Let[4]

$$N_\vartheta(t, x_1, x_2, x_3) := \mathrm{trunk}\left(\frac{cM_0(t, x_1, x_2, x_3)}{h}\right),$$

$$N_\varpi(t, x_1, x_2, x_3) := \mathrm{trunk}\left(\frac{cM_4(t, x_1, x_2, x_3)}{h}\right).$$

Hence, functions $N_\vartheta(t, x_1, x_2, x_3)$ and $N_\varpi(t, x_1, x_2, x_3)$ have integer values.
In that case with high precision:

$$\tilde{\varphi}(t, x_1, x_2, x_3, x_4, x_5) \approx \varphi(t, x_1, x_2, x_3) e^{i\left(x_5 \frac{h}{c} N_\vartheta(t,x_1,x_2,x_3) + x_4 \frac{h}{c} N_\varpi(t,x_1,x_2,x_3)\right)}$$

and Fourier series for $\tilde{\varphi}$ is of the following form:

$$\tilde{\varphi}(t, \vec{x}, x_4, x_5) = \varphi(t, \vec{x}) \sum_{n,s} \delta_{-n, N_\vartheta\langle t, \vec{x}\rangle} \, \delta_{-s, N_\varpi\langle t, \vec{x}\rangle} e^{-i\frac{h}{c}(nx_5 + sx_4)}. \qquad (14)$$

The integer numbers $n$ and $s$ are called *mass numbers*.
From properties of $\delta$: in every point $\langle t, \vec{x}\rangle$: either $\tilde{\varphi}(t, \vec{x}, x_4, x_5) = 0$ or integer numbers $n_0$ and $s_0$ exist for which:

$$\tilde{\varphi}(t, \vec{x}, x_4, x_5) = \varphi(t, \vec{x}) e^{-i\frac{h}{c}(n_0 x_5 + s_0 x_4)}.$$

Here if $m_0 := (n_0^2 + s_0^2)^{0.5}$ and $m := (h/c)^2 m_0$ then $m$ is called *a mass* of $\tilde{\varphi}$.

That is for every space-time point: either this point is empty or single mass is placed in this point.

## QFT

Let [10] Á be some unitary space such that linear operators $\psi_s(\vec{x})$ ($s \in \{1,2,3,4\}$) act on all elements of this space. And let these operators satisfies to the following conditions:

1. Á contains an element $\tilde{F}_0$ such that: $\tilde{F}_0^\dagger \tilde{F}_0 = 1$ and $\psi_s \tilde{F}_0 = \tilde{0}$, hence, $\tilde{F}_0^\dagger \psi_s^\dagger = \tilde{0}^\dagger$;

2. $\{\psi_s(\vec{x}), \psi_{s'}(\vec{x})\} := \psi_s(\vec{x})\psi_{s'}(\vec{x}) + \psi_{s'}(\vec{x})\psi_s(\vec{x}) = \hat{0}$ and $\{\psi_s^\dagger(\vec{x}), \psi_{s'}^\dagger(\vec{x})\} = \hat{0}$;

---

[4] Function trunk ($x$) returns the integer part of a real number $x$ by removing the fractional part. For example:
trunk (2.0857) = 2.

3. Here exists an operator $\hat{1}$ (*identifier*) such that for every element $\tilde{F}$ of Á: $\hat{1}\tilde{F} = \tilde{F}$;

4. $\{\psi_{s'}^\dagger(\vec{y}), \psi_s(\vec{x})\} = \delta(\vec{y} - \vec{x})\delta_{s,s'}\hat{1}$.

Let and $\Omega$ ($\Omega \in R^3$) be a domain such that if $\vec{x} \in \Omega$ then $|x_r| < c\pi/h$ for $r \in \{1,2,3\}$.

Let $\Re_\Omega$ be a set of functions such that if element $\phi(\vec{x})$ of this set: if $\vec{x} \notin \Omega$ then $\phi(\vec{x}) = 0$.

Hence,

$$\int_\Omega d^3\vec{x} \cdot \phi(\vec{x}) = \int_{-\frac{c\pi}{h}}^{\frac{c\pi}{h}} dx_1 \int_{-\frac{c\pi}{h}}^{\frac{c\pi}{h}} dx_2 \int_{-\frac{c\pi}{h}}^{\frac{c\pi}{h}} dx_3 \cdot \phi(\vec{x})$$

and for each element $\phi(\vec{x})$ of $\Re_\Omega$ exists a number $J_\phi$ such that

$$J_\phi = \int_{(\Omega)} d^3\vec{x} \cdot \phi^*(\vec{x})\phi(\vec{x}).$$

Therefore, $\Re_\Omega$ is an unitary space with the following scalar product:

$$\tilde{u} * \tilde{v} := \int_{(\Omega)} d^3\vec{x} \cdot \tilde{u}^*(\vec{x})\tilde{v}(\vec{x}).$$

This space has an orthonormalized basis with the following elements:

$$\varsigma_{w,\vec{p}}(\vec{x}) := \begin{cases} \left(\frac{h}{2\pi c}\right)^{\frac{3}{2}} e^{-i\frac{h}{c}\vec{p}\cdot\vec{x}}, & \text{if } -\frac{\pi c}{h} \leq x_k \leq \frac{\pi c}{h}, \\ 0, & \text{otherwize.} \end{cases}$$

with $k \in \{1,2,3\}$ and with natural $p_1, p_2, p_3$ (here: $\vec{p} := \langle p_1, p_2, p_3 \rangle$ and $\vec{p} \cdot \vec{x} := p_1 x_1 + p_2 x_2 + p_3 x_3$).

In that case if

$$\Psi(t, \vec{x}) := \sum_{s=1}^{4} \varphi_s(t, \vec{x}) \psi_s^\dagger(\vec{x}) \tilde{F}_0$$

then [11]

$$\int_{(\Omega)} d^3\vec{x}' \cdot \Psi^\dagger(t, \vec{x}')\Psi(t, \vec{x}) = \rho_A(t, \vec{x}).$$

Denote:

$$\hat{N}_a(\vec{y}) := \psi_a^\dagger(\vec{y})\psi_a(\vec{y}).$$

In that case the average value of operator $\hat{N}_a(\vec{y})$ is the following:

$$\langle\hat{N}_a(\vec{y})\rangle_\Psi := \int_{(\Omega)} d^3\vec{x} \cdot \hat{N}_a(\vec{y})\rho_A(t, \vec{x}).$$

Therefore [12],

$$\langle\hat{N}_a(\vec{y})\rangle_\Psi = \varphi_a^*\varphi_a.$$

That is operator $\hat{N}_a(\vec{y})$ brings the *a*-component of the event A probability density.

If

$$\Psi_a(t,\vec{x}) := \psi_a(\vec{y})\Psi(t,\vec{x})$$

then [13]

$$\langle \widehat{N}_a(\vec{y})\rangle_{\Psi_a} = 0.$$

Therefore, $\psi_a$ "annihilates" the *a*-component of the event *A* probability density.

If

$$\Psi_{a+}(t,\vec{x}) := \psi_a^\dagger(\vec{y})\Psi(t,\vec{x})$$

then

$$\langle \widehat{N}_a(\vec{y})\rangle_{\Psi_{a+}} = 1 - \varphi_a^*(\vec{y})\varphi_a(\vec{y}).$$

That is if $\varphi_a(\vec{y}) = 0$ then $\langle \widehat{N}_a(\vec{y})\rangle_{\Psi_{a+}} = 1$ and if $\varphi_a^*(\vec{y})\varphi_a(\vec{y}) = 1$ then $\langle \widehat{N}_a(\vec{y})\rangle_{\Psi_{a+}} = 0$ (If *a*-component is not contained in the initial state then operator $\psi_a^\dagger$ adds it. And if *a*-component is contained in the initial state then operator $\psi_a^\dagger$ turns everything to zero). Therefore, $\psi_a^\dagger$ "creates" the *a*-component of the event *A* probability density if this component absented.

I call operator $\psi_a^\dagger(\vec{y})$ *an creating operator* and $\psi_a(\vec{y})$ *an annihilation operator of probability of event A in point* $\vec{y}$. Operator $\psi_a^\dagger(\vec{y})$ is not an operator of a particle creating in point $\vec{y}$, but this operator varieties a probability of event *A* in that point. An operator $\psi_a(\vec{y})$ annihilates this probability in this point.

Similar operators are fundamental notions of the Quantum Fields Theory, originated by Fock[5], Pauli[6], Heisenberg[7], Bethe[8], Tomonaga[9], Schwinger[10], Feynman[11], Dyson[12]

Let $N_\varpi(t, x_1, x_2, x_3) = 0$ and $N_\vartheta(t, x_1, x_2, x_3) = n_0$. In that case from (14):

$$\tilde{\varphi}(t,\vec{x},x_4,x_5) = \varphi(t,\vec{x})e^{-i\frac{h}{c}(n_0 x_5)}.$$

Let $\mathfrak{J}_0$ be a space spanned of sub basis with the following elements:

$$\frac{h}{2\pi c}e^{-i\frac{h}{c}(n_0 x_5)}\varepsilon_k$$

That is $\mathfrak{J}_0$ is a subspace of unitary space $\mathfrak{J}$.

On $\mathfrak{J}_0$ the Hamiltonian (10) has the following shape in the accordance with equation (12):

---

[5] Vladimir Aleksandrovich Fock, 1898-1974
[6] Wolfgang Pauli, 1900-1958
[7] Werner Karl Heisenberg, 1901-1976
[8] Hans Albrecht Bethe, 1906-2005
[9] Shinichiro Tomonaga, 1906-1979
[10] Julian Seymour Schwinger, 1918-1994
[11] Richard Phillips Feynman, 1918-1988
[12] Freeman John Dyson, brn. 1923

$$\widehat{H}_l := c \sum_{k=1}^{3} \left( \beta^{[k]} i \partial_k + \frac{h}{c} n_0 \gamma^{[0]} + \widehat{G} \right)$$

with

$$\widehat{G} := \frac{1}{c} \sum_{\mu=0}^{3} \beta^{[\mu]} \left( i \partial_\mu + F_\mu + 0.5 g_1 Y B_\mu \right).$$

Denote:

$$\widehat{H}_0 := c \sum_{k=1}^{3} \left( \beta^{[k]} i \partial_k + \frac{h}{c} n_0 \gamma^{[0]} \right).$$

Denote [14]:

$$\omega(\vec{k}) := \sqrt{\vec{k}^2 + n_0^2} = \sqrt{k_1^2 + k_2^2 + k_3^2 + n_0^2},$$

$$e_1(\vec{k}) := \frac{1}{2\omega(\vec{k})(\omega(\vec{k}) + n_0)} \begin{bmatrix} \omega(\vec{k}) + n_0 + k_3 \\ k_1 + ik_2 \\ \omega(\vec{k}) + n_0 - k_3 \\ -k_1 - ik_2 \end{bmatrix},$$

$$e_2(\vec{k}) := \frac{1}{2\omega(\vec{k})(\omega(\vec{k}) + n_0)} \begin{bmatrix} k_1 - ik_2 \\ \omega(\vec{k}) + n_0 - k_3 \\ -k_1 + ik_2 \\ \omega(\vec{k}) + n_0 + k_3 \end{bmatrix},$$

$$e_3(\vec{k}) := \frac{1}{2\omega(\vec{k})(\omega(\vec{k}) + n_0)} \begin{bmatrix} -\omega(\vec{k}) - n_0 + k_3 \\ k_1 + ik_2 \\ \omega(\vec{k}) + n_0 + k_3 \\ k_1 + ik_2 \end{bmatrix},$$

$$e_4(\vec{k}) := \frac{1}{2\omega(\vec{k})(\omega(\vec{k}) + n_0)} \begin{bmatrix} k_1 - ik_2 \\ -\omega(\vec{k}) - n_0 - k_3 \\ k_1 - ik_2 \\ \omega(\vec{k}) + n_0 - k_3 \end{bmatrix}.$$

In that case:

$$\widehat{H}_0 e_1(\vec{k}) \left( \frac{h}{2\pi c} \right)^{\frac{3}{2}} e^{-i\frac{h}{c}(\vec{k}\vec{x})} = h\omega(\vec{k}) e_1(\vec{k}) \left( \frac{h}{2\pi c} \right)^{\frac{3}{2}} e^{-i\frac{h}{c}(\vec{k}\vec{x})}.$$

Hence, function $e_1(\vec{k}) \left( \frac{h}{2\pi c} \right)^{\frac{3}{2}} e^{-i\frac{h}{c}(\vec{k}\vec{x})}$ is an eigenvector of $\widehat{H}_0$ with eigenvalue $(h\omega(\vec{k}))$.

And function $e_2(\vec{k})\left(\frac{h}{2\pi c}\right)^{\frac{3}{2}} e^{-i\frac{h}{c}(\vec{k}\vec{x})}$ too. But functions $e_3(\vec{k})\left(\frac{h}{2\pi c}\right)^{\frac{3}{2}} e^{-i\frac{h}{c}(\vec{k}\vec{x})}$ and $e_4(\vec{k})\left(\frac{h}{2\pi c}\right)^{\frac{3}{2}} e^{-i\frac{h}{c}(\vec{k}\vec{x})}$ are eigenvectors of $\widehat{H}_0$ with eigenvalue $(-h\omega(\vec{k}))$.

Let

$$b_{r,\vec{k}} := \left(\frac{h}{2\pi c}\right)^3 \sum_{j'=1}^{4} \int_{(\Omega)} d^3\vec{x}' \cdot e^{-i\frac{h}{c}(\vec{k}\vec{x}')} e^*_{r,j'}(\vec{k})\psi_{j'}(\vec{x}').$$

In that case [15]:

$$\psi_j(\vec{x}) = \sum_{\vec{k}} e^{-i\frac{h}{c}(\vec{k}\vec{x})} \sum_{r=1}^{4} b_{r,\vec{k}}\, e_{r,j}(\vec{k}).$$

$$\{b^\dagger_{s,\vec{k}'}, b_{r,\vec{k}}\} = \left(\frac{h}{2\pi c}\right)^3 \delta_{s,r}\delta_{\vec{k},\vec{k}'},$$

$$\{b^\dagger_{s,\vec{k}'}, b^\dagger_{r,\vec{k}}\} = 0 = \{b_{s,\vec{k}'}, b_{r,\vec{k}}\},$$

$$b_{r,\vec{k}}\tilde{F}_0 = \tilde{0}.$$

An operator $\mathcal{H}$ such that:

$$\mathcal{H}(\vec{x}) := \sum_{s=1}^{4} \psi^\dagger_s(\vec{x}) \sum_{k=1}^{4} \widehat{H}_{s,k}\psi_k(\vec{x})$$

is called *Hamiltonian density* of $\widehat{H}$.

This operator fulfils to the following condition [16]:

$$\int_{(\Omega)} d^3\vec{x} \cdot \mathcal{H}(\vec{x})\Psi(t,\vec{y}) = i\partial_t \Psi(t,\vec{y}).$$

Hence, a Hamiltonian density defined a variation of the event $A$ probability in a space-time point $\vec{y}$ on the time instant $t$.

If

$$\mathcal{H}_0(\vec{x}) := \sum_{s=1}^{4} \psi^\dagger_s(\vec{x}) \sum_{k=1}^{4} \widehat{H}_{0,s,k}\psi_k(\vec{x})$$

then [17]

$$\int_{(\Omega)} d^3\vec{x} \cdot \mathcal{H}_0(\vec{x}) = \left(\frac{h}{2\pi c}\right)^3 \sum_{\vec{k}} h\omega(\vec{k})\left(\sum_{\alpha=1}^{2} b^\dagger_{\alpha,\vec{k}} b_{\alpha,\vec{k}} - \sum_{\alpha=3}^{4} b^\dagger_{\alpha,\vec{k}} b_{\alpha,\vec{k}}\right).$$

If

$$c_r(t,\vec{p}) := \left(\frac{h}{2\pi c}\right)^3 \sum_{j'=1}^{4} \int_{(\Omega)} d^3\vec{x}' \cdot e^*_{r,j'}(\vec{p})\varphi_{j'}(t,\vec{x}')e^{-i\frac{h}{c}(\vec{p}\vec{x}')}$$

then [18]

$$\int_{(\Omega)} d^3\vec{x} \cdot \Psi(t,\vec{x}) = \left(\frac{h}{2\pi c}\right)^3 \sum_{\vec{p}} \sum_{r=1}^{4} c_r(t,\vec{p}) \, b^\dagger_{r,\vec{p}} \tilde{F}_0.$$

Hence, if

$$\tilde{\Psi}(t,\vec{p}) := \left(\frac{h}{2\pi c}\right)^3 \sum_{r=1}^{4} c_r(t,\vec{p}) \, b^\dagger_{r,\vec{p}} \tilde{F}_0$$

then

$$\int_{(\Omega)} d^3\vec{x} \cdot \Psi(t,\vec{x}) = \sum_{\vec{p}} \tilde{\Psi}(t,\vec{p}).$$

Let

$$\hat{n}_{s,\vec{q}} := b^\dagger_{s,\vec{q}} b_{s,\vec{q}}$$

and let

$$\langle \hat{n}_{s,\vec{q}} \rangle_{\tilde{\Psi}} := \sum_{\vec{p}'} \sum_{\vec{p}} \tilde{\Psi}^\dagger(t,\vec{p}') \hat{n}_{s,\vec{q}} \tilde{\Psi}(t,\vec{p})$$

then [19]

$$\langle \hat{n}_{s,\vec{q}} \rangle_{\tilde{\Psi}} = c^*_s(t,\vec{q}) c_s(t,\vec{q}) = |c_s(t,\vec{q})|^2.$$

Therefore, operator $\hat{n}_{s,\vec{q}}$ brings the *s*-component of $\tilde{\Psi}$ similar to $\langle \widehat{N}_a(\vec{y}) \rangle_\Psi$. Operator $b_{s,\vec{q}}$ "annihilates" this component similar to how $\psi_a$ "annihilates" the *a*-component of the event *A* probability density. Operator $b_{r,\vec{p}}$ is called *operator of particle $\langle r,\vec{p}\rangle$-state annihilation*, and operator $b^\dagger_{r,\vec{p}}$ is called *operator of a particle $\langle r,\vec{p}\rangle$-state creation*.

If

$$E_{0,\tilde{\Psi}} := \sum_{\vec{p}'} \sum_{\vec{p}} \tilde{\Psi}^\dagger(t,\vec{p}') \left( \int_{(\Omega)} d^3\vec{x} \cdot \mathcal{H}_0(\vec{x}) \right) \tilde{\Psi}(t,\vec{p})$$

then $E_{0,\tilde{\Psi}}$ is called *average value of energy of free state* $\Psi$ and [20]

$$E_{0,\tilde{\Psi}} = \left(\frac{h}{2\pi c}\right)^3 \sum_{\vec{k}} h\omega(\vec{k}) \left( \sum_{\alpha=1}^{2} |c_s(t,\vec{q})|^2 - \sum_{\alpha=3}^{4} |c_s(t,\vec{q})|^2 \right).$$

That value is not positive defined.

But our physics devices cannot recognize of energy negative or positive sign. Negative energy is registered the same way as positive energy. This misunderstanding was eliminated in 1930 by P. A. M. Dirac which introduced into practice notion of *antiparticle*. In 1932 Carl D. Anderson[13] registered tracks of cosmic particles which warped under magnetic field like electron tracks but in the opposite direction.

Mathematics of it is the following [21]: let

$$\begin{cases} v_{(1)}(\vec{k}) := \gamma^{[0]} e_3(\vec{k}), \\ v_{(2)}(\vec{k}) := \gamma^{[0]} e_4(\vec{k}), \\ e_{(1)}(\vec{k}) := e_1(\vec{k}), \\ e_{(2)}(\vec{k}) := e_2(\vec{k}). \end{cases}$$

and let:

$$\begin{cases} d_{1,\vec{k}} := b^\dagger_{3,-\vec{k}}, \\ d_{2,\vec{k}} := b^\dagger_{4,-\vec{k}} \end{cases}$$

and [22]

$$\widehat{H}^{[0]} := \left(\frac{h}{2\pi c}\right)^3 \int_{(\Omega)} d^3\vec{x} \cdot \mathcal{H}_0(\vec{x}) + h \sum_{\vec{k}=-\vec{L}}^{\vec{L}} \omega(\vec{k}) \hat{1}.$$

In that case:

$$\widehat{H}^{[0]} := \left(\frac{h}{2\pi c}\right)^3 \sum_{\vec{k}=-\vec{L}}^{\vec{L}} h\omega(\vec{k}) \sum_{\alpha=1}^{2} \left(b^\dagger_{\alpha,\vec{k}} b_{\alpha,\vec{k}} + d^\dagger_{\alpha,\vec{k}} d_{\alpha,\vec{k}}\right).$$

This Hamiltonian is positive defined.

Here $b^\dagger_{\alpha,\vec{k}}$ are *creation operators*, and $b_{\alpha,\vec{k}}$ are *annihilation operators* of *n-lepton* with *momentum* $\vec{k}$ and *spin index* α; $d^\dagger_{\alpha,\vec{k}}$ are *creation operators*, and $d_{\alpha,\vec{k}}$ are *annihilation operators* of *anti-n-lepton* with *momentum* $k$ and *spin index* α.

Therefore, principal concepts of Quantum Field Theory are deduced from properties of physics events probabilities.

---

[13] Carl David Anderson, 1905-1991

## Electroweak equations

Let us consider [23, 24] the subspace $\mathfrak{J}_{ev}$ of the space $\mathfrak{J}$ spanned of the following subbasis [25]:

$$\mathfrak{J}_{ev} := \left\{ \begin{array}{c} \frac{h}{2\pi c}\sqrt{\frac{2\pi n_0}{\sinh(2\pi n_0)}}\left(\cosh\left(\frac{h}{c}n_0 x_4\right) + \sinh\left(\frac{h}{c}n_0 x_4\right)\right)\varepsilon_1, \\ \frac{h}{2\pi c}\sqrt{\frac{2\pi n_0}{\sinh(2\pi n_0)}}\left(\cosh\left(\frac{h}{c}n_0 x_4\right) + \sinh\left(\frac{h}{c}n_0 x_4\right)\right)\varepsilon_2, \\ \frac{h}{2\pi c}\sqrt{\frac{2\pi n_0}{\sinh(2\pi n_0)}}\left(\cosh\left(\frac{h}{c}n_0 x_4\right) + \sinh\left(\frac{h}{c}n_0 x_4\right)\right)\varepsilon_3, \\ \frac{h}{2\pi c}\sqrt{\frac{2\pi n_0}{\sinh(2\pi n_0)}}\left(\cosh\left(\frac{h}{c}n_0 x_4\right) + \sinh\left(\frac{h}{c}n_0 x_4\right)\right)\varepsilon_4, \\ \frac{h}{2\pi c}e^{-i\frac{h}{c}n_0 x_5}\varepsilon_1, \frac{h}{2\pi c}e^{-i\frac{h}{c}n_0 x_5}\varepsilon_2, \\ \frac{h}{2\pi c}e^{-i\frac{h}{c}n_0 x_5}\varepsilon_3, \frac{h}{2\pi c}e^{-i\frac{h}{c}n_0 x_5}\varepsilon_4, \end{array} \right\}$$

with some integer numbers $n_0$.

Let $U$ be any linear transformation of space $\mathfrak{J}_{ev}$ such that for every $\tilde{\varphi}$: if $\tilde{\varphi} \in \mathfrak{J}_{ev}$ then

$$\begin{cases} (U\tilde{\varphi}, U\tilde{\varphi}) = \rho_A, \\ (U\tilde{\varphi}, \beta^{[k]}U\tilde{\varphi}) = -\frac{j_{A,k}}{c}. \end{cases}$$

In that case matrix $U$ can be represented by the product of matrices of the following shape:

$\tilde{U}(\chi)$, [14]

$$U^{(+)} := \begin{bmatrix} 1_2 & 0_2 & 0_2 & 0_2 \\ 0_2 & (u+iv)1_2 & 0_2 & (k+is)1_2 \\ 0_2 & 0_2 & 1_2 & 0_2 \\ 0_2 & (-k+is)1_2 & 0_2 & (u-iv)1_2 \end{bmatrix}, \quad [15]$$

$$U^{(-)} := \begin{bmatrix} (a+ib)1_2 & 0_2 & (c+iq)1_2 & 0_2 \\ 0_2 & 1_2 & 0_2 & 0_2 \\ (-c+iq)1_2 & 0_2 & (a-ib)1_2 & 0_2 \\ 0_2 & 0_2 & 0_2 & 1_2 \end{bmatrix}$$

here $u(t, \vec{k})$, $v(t, \vec{k})$, $k(t, \vec{k})$, $s(t, \vec{k})$, $a(t, \vec{k})$, $b(t, \vec{k})$, $c(t, \vec{k})$, $q(t, \vec{k})$ are real functions such that

$u^2 + v^2 + k^2 + s^2 = 1$, $a^2 + b^2 + c^2 + q^2 = 1$.

Denote:

---

[14] see formula (13)
[15] This matrix concerns to antiparticles (see part QFT of this article).

$$\ell_o := \frac{1}{2\sqrt{1-a^2}} \begin{bmatrix} \left(b + \sqrt{1-a^2}\right) 1_4 & (q - ic) 1_4 \\ (q + ic) 1_4 & \left(-b + \sqrt{1-a^2}\right) 1_4 \end{bmatrix},$$

$$\ell_x := \frac{1}{2\sqrt{1-a^2}} \begin{bmatrix} \left(-b + \sqrt{1-a^2}\right) 1_4 & (-q + ic) 1_4 \\ (-q - ic) 1_4 & \left(b + \sqrt{1-a^2}\right) 1_4 \end{bmatrix},$$

$\tilde{\varphi}_o := \ell_o \tilde{\varphi}, \tilde{\varphi}_x := \ell_o \tilde{\varphi}$

In that case:

$\tilde{\varphi}_o + \tilde{\varphi}_x = \tilde{\varphi},$

$$\begin{cases} U^{(-)\dagger} \gamma^{[0]} U^{(-)} \tilde{\varphi}_o = \left(a\gamma^{[0]} - \sqrt{1-a^2}\beta^{[4]}\right) \tilde{\varphi}_o, \\ U^{(-)\dagger} \beta^{[4]} U^{(-)} \tilde{\varphi}_o = \left(a\beta^{[4]} + \sqrt{1-a^2}\gamma^{[0]}\right) \tilde{\varphi}_o; \end{cases}$$

$$\begin{cases} U^{(-)\dagger} \gamma^{[0]} U^{(-)} \tilde{\varphi}_x = \left(a\gamma^{[0]} + \sqrt{1-a^2}\beta^{[4]}\right) \tilde{\varphi}_x, \\ U^{(-)\dagger} \beta^{[4]} U^{(-)} \tilde{\varphi}_x = \left(a\beta^{[4]} - \sqrt{1-a^2}\gamma^{[0]}\right) \tilde{\varphi}_x. \end{cases}$$

In space $\mathfrak{I}_{ev}$ equation (12) is equivalent to the following [26]:

$$\sum_{\mu=0}^{3} \beta^{[\mu]} \left(i\partial_\mu - e\hat{A}_\mu + 0.5(\hat{Z}_\mu + \widehat{W}_\mu)\right) \tilde{\varphi} + \left(-\gamma^{[0]} i\partial_5 - \beta^{[4]} i\partial_4\right) \tilde{\varphi} = 0.$$

with [27] [28]

$$\tilde{\varphi} := \begin{bmatrix} \varphi_{v,1} \\ \varphi_{v,2} \\ 0 \\ 0 \\ \varphi_{e,L,1} \\ \varphi_{e,L,2} \\ \varphi_{e,R,1} \\ \varphi_{e,R,2} \end{bmatrix}, \hat{A}_\mu := A_\mu \begin{bmatrix} 0_2 & 0_2 & 0_2 & 0_2 \\ 0_2 & 1_2 & 0_2 & 0_2 \\ 0_2 & 0_2 & 1_2 & 0_2 \\ 0_2 & 0_2 & 0_2 & 1_2 \end{bmatrix}..$$

$$\hat{Z}_\mu := Z_\mu \frac{1}{\sqrt{g_1^2 + g_2^2}} \begin{bmatrix} (g_1^2 + g_2^2) 1_2 & 0_2 & 0_2 & 0_2 \\ 0_2 & 2g_1^2 1_2 & 0_2 & 0_2 \\ 0_2 & 0_2 & (g_2^2 - g_1^2) 1_2 & 0_2 \\ 0_2 & 0_2 & 0_2 & 2g_1^2 1_2 \end{bmatrix},$$

$$\widehat{W}_\mu := g_2 \begin{bmatrix} 0_2 & 0_2 & (W_{1,\mu} - iW_{2,\mu}) 1_2 & 0_2 \\ 0_2 & 0_2 & 0_2 & 0_2 \\ (W_{1,\mu} + iW_{2,\mu}) 1_2 & 0_2 & 0_2 & 0_2 \\ 0_2 & 0_2 & 0_2 & 0_2 \end{bmatrix}.$$

here: $A_\mu$, $Z_\mu$, $W_{k,\mu}$ are real functions; $g_1$ and $g_2$ are real positive constants,

$$e := \frac{g_1 g_2}{\sqrt{g_1^2 + g_2^2}}.$$

Fields $W_{k,\mu}$ obey to the following equation [29]:

$$\left(-\frac{1}{c}\partial_t^2 + \sum_{s=1}^3 \partial_s^2\right) W_{k,\mu} = g_2^2 \left(\widetilde{W}_0^2 - \sum_{s=1}^3 \widetilde{W}_s^2\right) W_{k,\mu} + \Lambda. \tag{15}$$

Here:

$$\widetilde{W}_\nu := \begin{bmatrix} W_{0,\nu} \\ W_{1,\nu} \\ W_{2,\nu} \end{bmatrix} \text{ with } W_{0,\nu} := Z_\nu \frac{g_2}{\sqrt{g_1^2+g_2^2}} + A_\nu \frac{g_1}{\sqrt{g_1^2+g_2^2}}$$

and $\Lambda$ is the action of other components of field $W$ on $W_{k,\mu}$.

Equation (15) looks like the Klein-Gordon equation of field $W_{k,\mu}$ with mass

$$m := \frac{h}{c} g_2 \sqrt{\widetilde{W}_0^2 - \sum_{s=1}^3 \widetilde{W}_s^2}$$

and with additional terms of the $W_{k,\mu}$ interactions with other components of $W$.

This "mass" is invariant under the Lorentz transformations:

$$\widetilde{W}_0 \to \widetilde{W}_0' = \frac{\widetilde{W}_0 - \frac{v}{c}\widetilde{W}_k}{\sqrt{1-\left(\frac{v}{c}\right)^2}}, \widetilde{W}_k \to \widetilde{W}_k' = \frac{\widetilde{W}_k - \frac{v}{c}\widetilde{W}_0}{\sqrt{1-\left(\frac{v}{c}\right)^2}}, \widetilde{W}_j \to \widetilde{W}_j' = \widetilde{W}_j \text{ if } j \neq k,$$

is invariant under the turns of the $\langle \widetilde{W}_1, \widetilde{W}_2, \widetilde{W}_3 \rangle$ space:

$$\begin{cases} \widetilde{W}_k \to \widetilde{W}_k' = \widetilde{W}_k \cos\lambda - \widetilde{W}_s \sin\lambda, \\ \widetilde{W}_s \to \widetilde{W}_s' = \widetilde{W}_s \cos\lambda + \widetilde{W}_k \sin\lambda, \end{cases}$$

and invariant under a global weak isospin transformation $U^{(-)}$:

$$W_\nu \to W_\nu' = U^{(-)} W_\nu U^{(-)\dagger}$$

but is not invariant for a local transformation $U^{(-)}$. But local transformations for $W_{0,\mu}$, $W_{1,\mu}$, $W_{2,\mu}$ are insignificant since all three particles are very short-lived and a measurement of masses of these particles is practically possible only at the point $\langle t \approx 0, \vec{x} \approx \vec{0}\rangle$.

If

$$\alpha := \tan^{-1}\frac{g_1}{g_2}$$

then masses of $Z$ and $W$ fulfill the following ratio:

$$m_W = \frac{m_Z}{\cos \alpha}.$$

Therefore, the Glashow's electroweak theory without Higgs is deduced from properties of physics events probabilities.

## Chromatic oscillations

I call the following part of (8):

$$\sum_{k=0}^{3} \beta^{[k]} (\partial_k - i\Theta_k - i\Upsilon_k \gamma^{[5]})$$
$$+ \left( -iM_{\varsigma,0} \gamma_\varsigma^{[0]} + iM_{\varsigma,4} \varsigma^{[4]} - iM_{\eta,0} \gamma_\eta^{[0]} - iM_{\eta,4} \eta^{[4]} + iM_{\theta,0} \gamma_\theta^{[0]} + iM_{\theta,4} \theta^{[4]} \right) \varphi$$
$$= 0 \qquad (16)$$

*chromatic moving equation*.

The mass members of this equation form the following matrix sum:

$$\widehat{M} := \left( -iM_{\varsigma,0} \gamma_\varsigma^{[0]} + iM_{\varsigma,4} \varsigma^{[4]} - iM_{\eta,0} \gamma_\eta^{[0]} - iM_{\eta,4} \eta^{[4]} + iM_{\theta,0} \gamma_\theta^{[0]} + iM_{\theta,4} \theta^{[4]} \right) =$$

$$= \begin{bmatrix} 0 & 0 & -M_{\theta,0} & M_{\varsigma,0} - iM_{\eta,0} \\ 0 & 0 & M_{\varsigma,0} + iM_{\eta,0} & M_{\theta,0} \\ -M_{\theta,0} & M_{\varsigma,0} - iM_{\eta,0} & 0 & 0 \\ M_{\varsigma,0} + iM_{\eta,0} & M_{\theta,0} & 0 & 0 \end{bmatrix} +$$

$$+ i \begin{bmatrix} 0 & 0 & -M_{\theta,4} & M_{\varsigma,4} + iM_{\eta,4} \\ 0 & 0 & M_{\varsigma,4} - iM_{\eta,4} & M_{\theta,4} \\ M_{\theta,4} & -M_{\varsigma,4} - iM_{\eta,4} & 0 & 0 \\ -M_{\varsigma,4} + iM_{\eta,4} & -M_{\theta,4} & 0 & 0 \end{bmatrix},$$

Elements of these matrices can be rotated by the following octad elements [30]
$$\mathring{U} := \{ U_{1,2}(\varsigma), U_{1,3}(\vartheta), U_{2,3}(\alpha), U_{0,1}(\sigma), U_{0,2}(\phi), U_{0,3}(\iota), \widetilde{U}(\chi), \widehat{U}(\kappa) \}$$
where $\varsigma(t, \vec{x})$, $\vartheta(t, \vec{x})$, $\alpha(t, \vec{x})$, $\sigma(t, \vec{x})$, $\phi(t, \vec{x})$, $\iota(t, \vec{x})$, $\chi(t, \vec{x})$, $\kappa(t, \vec{x})$ are any real functions.

For example, if

$$\widehat{M}' := \left( -iM'_{\varsigma,0} \gamma_\varsigma^{[0]} + iM'_{\varsigma,4} \varsigma^{[4]} - iM'_{\eta,0} \gamma_\eta^{[0]} - iM'_{\eta,4} \eta^{[4]} + iM'_{\theta,0} \gamma_\theta^{[0]} + iM'_{\theta,4} \theta^{[4]} \right) =$$

$$= U_{2,3}^\dagger \widehat{M} U_{2,3}$$

then

$$\begin{cases} M'_{\zeta,0} = M_{\zeta,0}, \\ M'_{\eta,0} = M_{\eta,0} \cos 2\alpha + M_{\theta,0} \sin 2\alpha, \\ M'_{\theta,0} = M_{\theta,0} \cos 2\alpha - M_{\eta,0} \sin 2\alpha, \\ M'_{\zeta,4} = M_{\zeta,4}, \\ M'_{\eta,4} = M_{\eta,4} \cos 2\alpha + M_{\theta,4} \sin 2\alpha, \\ M'_{\theta,4} = M_{\theta,4} \cos 2\alpha - M_{\eta,4} \sin 2\alpha. \end{cases}$$

Therefore, matrix $U_{2,3}$ makes an oscillation between green and blue colors. And this transformation of equation (16) bends time-space as the following:

$$\begin{cases} \dfrac{\partial x_2}{\partial x_2'} = \cos 2\alpha, \\ \dfrac{\partial x_3}{\partial x_2'} = -\sin 2\alpha, \\ \dfrac{\partial x_2}{\partial x_3'} = -\sin 2\alpha, \\ \dfrac{\partial x_3}{\partial x_3'} = \cos 2\alpha, \\ \dfrac{\partial x_0}{\partial x_2'} = \dfrac{\partial x_1}{\partial x_2'} = \dfrac{\partial x_0}{\partial x_3'} = \dfrac{\partial x_1}{\partial x_3'} = 0. \end{cases}$$

Therefore, the oscillation between blue and green colors bends the space in the $x_2$, $x_3$ directions.

One more example: if

$$\widehat{M}'' := \left( -iM''_{\zeta,0} \gamma_\zeta^{[0]} + iM''_{\zeta,4} \zeta^{[4]} - iM''_{\eta,0} \gamma_\eta^{[0]} - iM''_{\eta,4} \eta^{[4]} + iM''_{\theta,0} \gamma_\theta^{[0]} + iM''_{\theta,4} \theta^{[4]} \right) =$$

$$= U_{0,1}^{-1} \widehat{M} U_{0,1}$$

then

$$\begin{cases} M''_{\zeta,0} = M_{\zeta,0}, \\ M''_{\eta,0} = M_{\eta,0} \cosh 2\sigma - M_{\theta,4} \sinh 2\sigma, \\ M''_{\theta,0} = M_{\theta,0} \cosh 2\sigma + M_{\eta,4} \sinh 2\sigma, \\ M''_{\zeta,4} = M_{\zeta,4}, \\ M''_{\eta,4} = M_{\eta,4} \cosh 2\sigma + M_{\theta,0} \sinh 2\sigma, \\ M''_{\theta,4} = M_{\theta,4} \cosh 2\sigma - M_{\eta,0} \sinh 2\sigma. \end{cases}$$

Therefore, matrix $U_{0,1}$ makes an oscillation between green and blue colors with an oscillation between upper and lower mass members.. And this transformation of equation (16) bends time-space as the following:

$$\begin{cases} \dfrac{\partial x_1}{\partial x_1'} = \cosh 2\sigma, \\ \dfrac{\partial t}{\partial x_1'} = \dfrac{1}{c}\sinh 2\sigma, \\ \dfrac{\partial x_1}{\partial t'} = c\sinh 2\sigma, \\ \dfrac{\partial t}{\partial t'} = \cosh 2\sigma, \\ \dfrac{\partial x_2}{\partial t'} = \dfrac{\partial x_3}{\partial t'} = \dfrac{\partial x_2}{\partial x_1'} = \dfrac{\partial x_3}{\partial x_1'} = 0. \end{cases} \quad (18)$$

Therefore, the oscillation between blue and green colors with the oscillation between upper and lower mass members bends the space in the $t$, $x_1$ directions.

Such transformation with elements of set $\mathring{U}$ add to equation (16) gauge fields of the following shape: $U_{k,l}^{-1}(\xi)\partial_s U_{k,l}(\xi)$ where: $U_{k,l}(\xi) \in \mathring{U}$. And for every element $U_{k,l}(\xi)$ of $\mathring{U}$ exists [31] matrix $\Lambda_{k,l}$ such that

$$U_{k,l}^{-1}(\xi)\partial_s U_{k,l}(\xi) = \Lambda_{k,l}\partial_s(\xi)$$

and for every product $U$ of $\mathring{U}$'s elements real functions $G^r{}_s(t, x_1, x_2; x_3)$ exist such that

$$U^{-1}(\xi)\partial_s U(\xi) = \frac{g_3}{2}\sum_{r=1}^{8} \Lambda_r G_s^r$$

with some real constant $g_3$ (similar to 8 gluons).

From (18): the oscillation between upper and lower mass members bends the space in the $t$, $x_1$ directions with

$$\frac{\partial t}{\partial t'} = \cosh 2\sigma,$$
$$\frac{\partial x_1}{\partial t'} = c\sinh 2\sigma.$$

Hence, if $v$ is the velocity of a coordinate system $<t',x_1'>$ in the coordinate system $<t,x_1>$ then

$$\sinh 2\sigma = \frac{\left(\frac{v}{c}\right)}{\sqrt{1-\left(\frac{v}{c}\right)^2}}, \cosh 2\sigma = \frac{1}{\sqrt{1-\left(\frac{v}{c}\right)^2}}.$$

Therefore,

$$v = c\tanh 2\sigma.$$

Let

$$2\sigma := \omega(x_1)\frac{t}{x_1}$$

with

$$\omega(x_1) := \frac{\lambda}{|x_1|}.$$

where λ is a real constant with positive numerical value.

Fig. 1 shows the dependency of a system $<t',x_1'>$ velocity $v(t; x_1)$ on $x_1$ in system $<t,x_1>$.

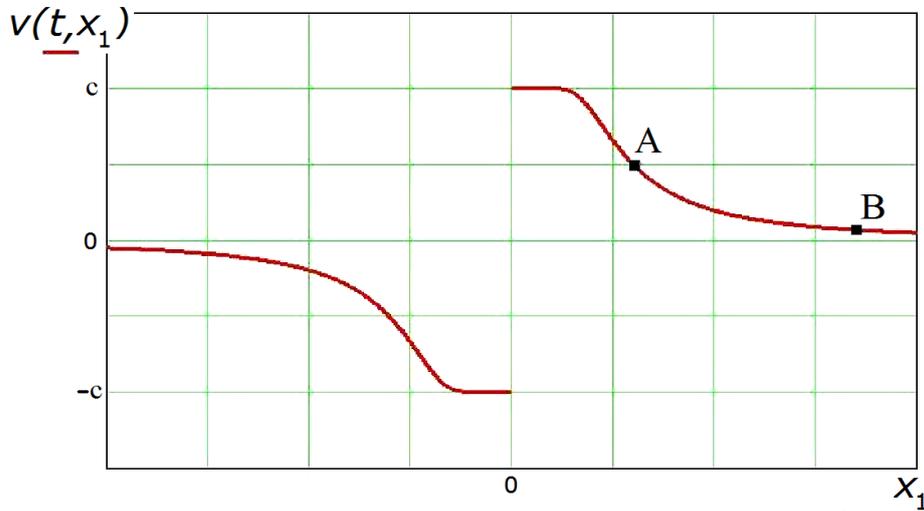

Fig.1

Let a black hole be placed in a point $O$ (Fig. 2). Then a tremendous number of quarks states oscillate in this point. These oscillations bend time-space and if $t$ has some fixed volume, $x_1 > 0$, and $\Lambda := \lambda t$ then

$$v(x_1) = c \tanh \frac{\Lambda}{x_1^2}.$$

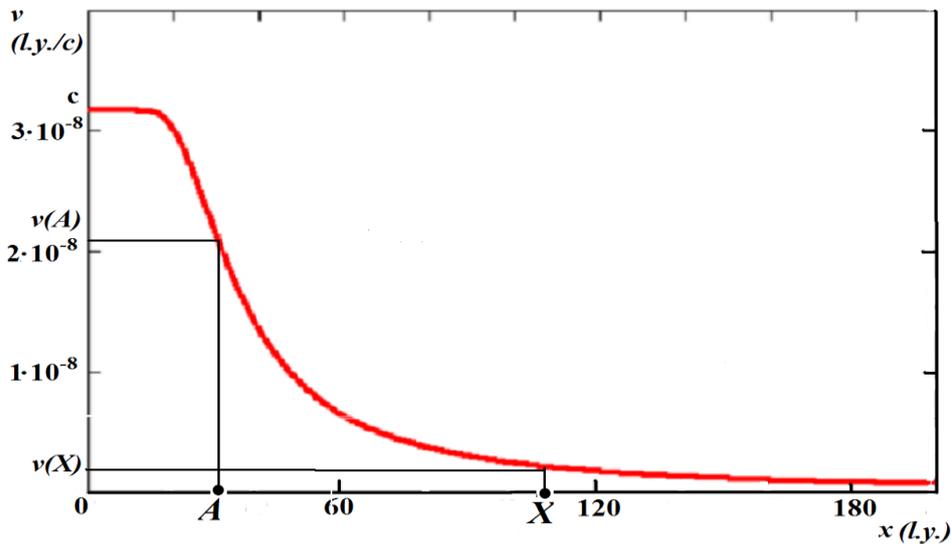

Fig.2

A dependency of $v(x_1)$ (light years/c.) on $x_1$ (light years) with $\Lambda = 741{:}907$ is shown in Fig. 2.

Let a placed in a point $A$ observer be stationary in the coordinate system $<t,x_1>$. Hence, in the coordinate system $<t',x_1'>$ this observer is flying to the left to the point $O$ with velocity $-v(x_A)$. And point $X$ is flying to the left to the point $O$ with velocity $-v(x_1)$.

Consequently, the observer $A$ sees that the point $X$ flies away from him to the right with velocity

$$V_A(x_1) := c\tanh\left(\frac{\Lambda}{x_A^2} - \frac{\Lambda}{x_1^2}\right).$$

in accordance with the relativistic rule of addition of velocities.

Let $r := x_1 - x_A$ (i.e. $r$ is distance from $A$ to $X$), and

$$V_A(x_1) := c\tanh\left(\frac{\Lambda}{x_A^2} - \frac{\Lambda}{(x_A+r)^2}\right).$$

In that case Fig. 3 demonstrates the dependence of $V_A(r)$ on $r$ with $x_A = 25 \times 10^3$ l.y.

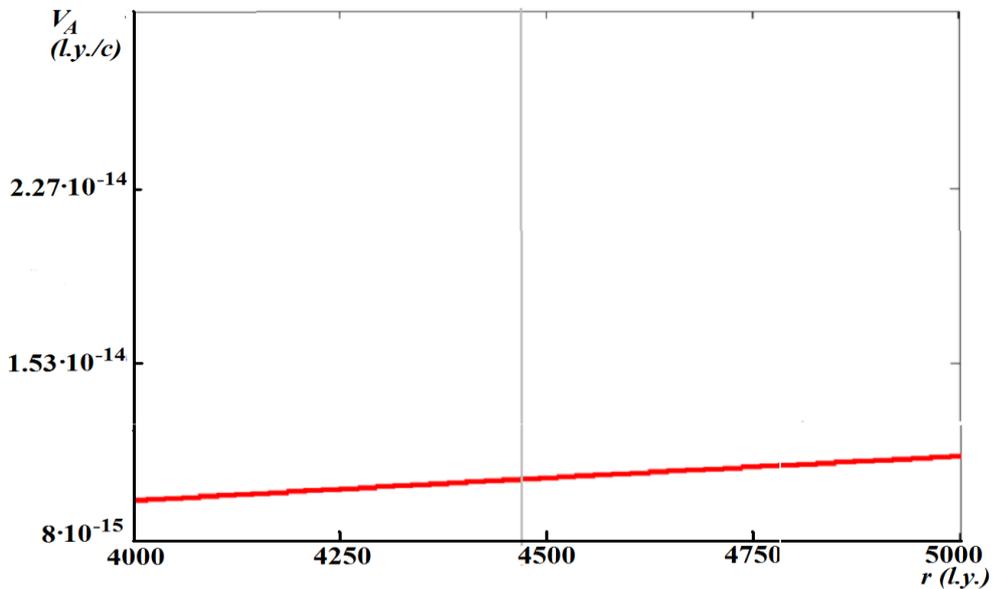

Fig.3

Hence, $X$ runs from $A$ with almost constant acceleration

$$\frac{V_A(r)}{r} =: H.$$

Fig. 4 demonstrates the dependence of $H$ on $r$ (the Hubble constant).

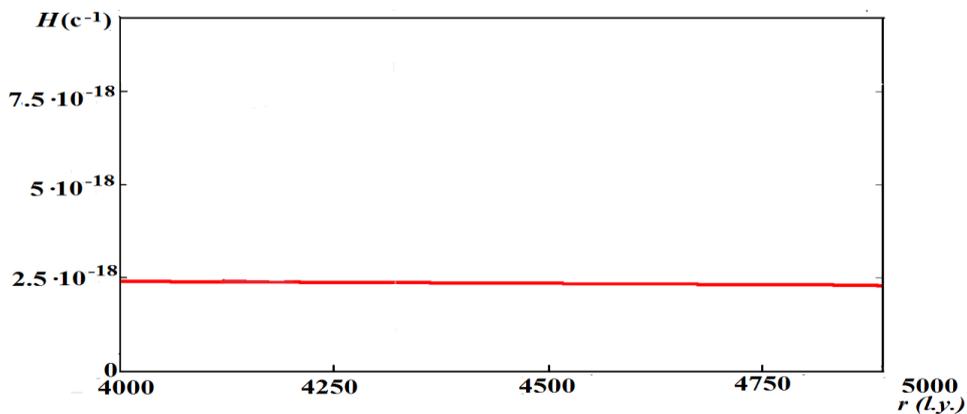

Fig.4

Therefore, accelerated expansion of the space is explained by oscillations $U_{0,1}$ of chromatic states. And oscillation $U_{1,2}$, $U_{1,3}$ and $U_{2,3}$ explain the discrepancy of quantity of the luminous matter in the space structures to the picture of gravitational interaction of stars in these structures.

If $g$ is an acceleration of system $<t',x_1'>$ as respects to system $<t,x_1>$ then

$$g(t,x_1) := \frac{\partial v}{\partial t} = \frac{\omega(x_1)}{x_1 \cosh\left(\omega(x_1)\frac{t}{x_1}\right)}.$$

Fig. 5 shows a dependency of this acceleration on $x_1$.

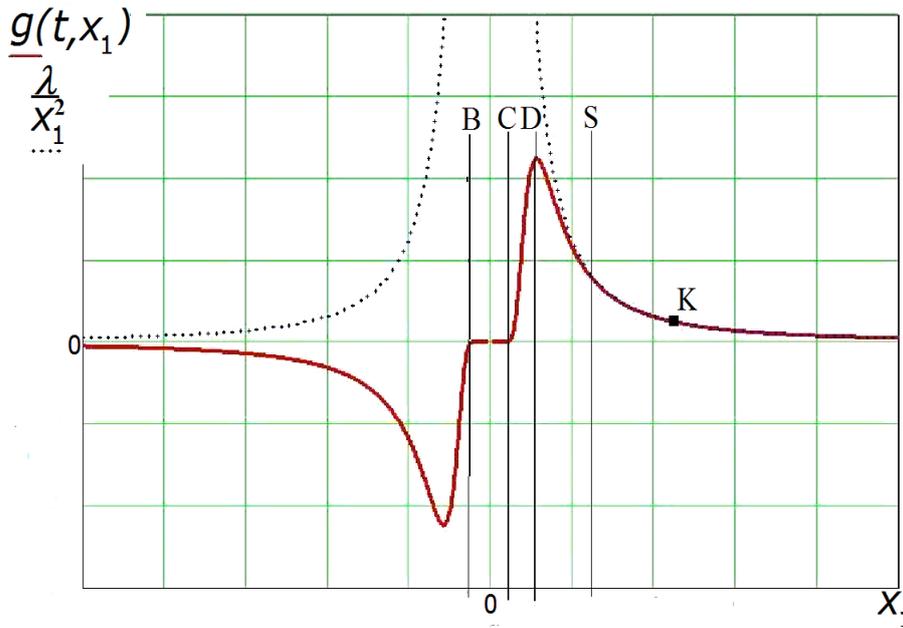

Fig.5

If an object immovable in system $<t,x_1>$ is placed in point K then in system $<t',x_1'>$ this object must move to the left with acceleration $g$ and

$$g(x_1) \approx \frac{\lambda}{x_1^2}.$$

I call: interval from S to $\infty$ *the Newton Gravity Zone*, interval from B to C *the Asymptotic Freedom Zone*, and interval from C to D *the Confinement Force Zone*.

## Conclusion

Therefore, all physical events are interpreted by the well-known fermions (leptons and quarks), gauge bosons ($A$, $W$, $Z$, $G$), their fields and gravity. Higgs, strings, Dark Energy and Dark Matter are not required.